# A novel energy resolved neutron imaging detector based on TPX3Cam for the CSNS

Jianqing Yang[1,2,3], Jianrong Zhou[2,3,4*], Xingfen Jiang[2,3,4], Jinhao Tan[2,3,5], Lianjun Zhang[2,3,5], Jianjin Zhou[2,3], Xiaojuan Zhou[2,3], Wenqin Yang[2,3,4], Yuanguang Xia[2,3], Jie Chen[2,3], XinLi Sun[1], Quanhu Zhang[1**], Zhijia Sun[2,3,4***], Yuanbo Chen[2,3,4]

1. Xi'an Research Institute of Hi-Tech, 710025, Xian, China.

2. Spallation Neutron Source Science Center, Dongguan, 523803, Guangdong, China;

3. State Key Laboratory of Particle Detection and Electronics, Institute of High Energy Physics, Chinese Academy of Sciences, Beijing, 100049, China;

4. University of Chinese Academy of Sciences, Beijing 100049, China

5. Harbin Engineering University, Harbin, Heilongjiang, China, 150000

**Abstract:** The China Spallation Neutron Source (CSNS) operates in pulsed mode and has a high neutron flux. This provides opportunities for energy resolved neutron imaging by using the TOF (Time Of Flight) approach. An Energy resolved neutron imaging instrument (ERNI) is being built at the CSNS but significant challenges for the detector persist because it simultaneously requires a spatial resolution of less than 100 μm, as well as a microsecond-scale timing resolution. This study constructs a prototype of an energy resolved neutron imaging detector based on the fast optical camera, TPX3Cam coupled with an image intensifier. To evaluate its performance, a series of proof of principle experiments were performed in the BL20 at the CSNS to measure the spatial resolution and the neutron wavelength spectrum, and perform neutron imaging with sliced wavelengths and Bragg edge imaging of the steel sample. A spatial resolution of 57 μm was obtained for neutron imaging by using the centroiding algorithm, the timing resolution was on the microsecond scale and the measured wavelength spectrum was identical to that measured by a beam monitor. In addition, any wavelengths can be selected for the neutron imaging of the given object, and the detector can be used for Bragg edge imaging. The results show that our detector has good performances and can satisfy the requirements of ERNI for detectors at the CSNS.

*Corresponding author : Jianrong Zhou, zhoujr@ihep.ac.cn, Dongguan, China

**Corresponding author : Quanhu Zhang, zhangqh102@sina.com, Xian, China

***Corresponding author : Zhijia Sun, sunzj@ihep.ac.cn, Beijing, China

# 1. Introduction

As a visible and non-destructive method of inspection, traditional neutron imaging evaluates the attenuation of a polychromatic neutron beam through a sample. However, the properties of the sample are integrated over the entire spectrum of the beam. Recent years have witnessed the development of energy resolved neutron imaging, which means that neutron imaging can be performed using a spectrum limited to a short energy band [1]. This is useful for analyzing energy dependence of the sample on neutrons through the contrast in neutron images of distinct energies, and the crystallographic information can be obtained using the Bragg edge imaging for crystal samples [2]. In principle, there are two ways of conducting energy resolved neutron imaging experiments. At the reactor sources, monochromator devices including velocity selectors [3] and double crystal monochromators [4] are available to select neutrons of specific energies. At the spallation sources, the TOF approach is a method for measuring the velocities of neutrons, i.e. energy or wavelength according to the de Broglie wave equation based on their flight distances as well as the TOF between the source pulse and the neutron arrival at the detector. Compared with reactor sources, the spallation sources have a significant advantage [5][6] because full energy scanning neutron imaging and Bragg edge imaging of the sample can be implemented in one experiment by using the TOF approach to improve the experimental efficiency. However, there are significant challenges for the detector in this context due to the requirements of high timing and spatial resolutions. At present, the spallation sources in operation include the JSNS (Japan Spallation Neutron Source) in Japan, the ISIS in the UK, the SNS (Spallation Neutron Source) in USA and the CSNS in China. The energy resolved neutron imaging system, RADEN has been constructed at the JSNS of J-PARC. The detectors applied to it include the μNID, nGEM and the LiTA12 [7] whose timing resolutions are on the dozens or hundreds of ns scale. However, their spatial resolutions are over 100μm. The IMAT imaging instrument has been installed in the ISIS [8]. The conceptual design of the energy resolved neutron imaging instrument, VENUS has been completed at the SNS and it is being constructed [9]. The MCP neutron counting detector based on Timepix3 chip readout has been chosen as the imaging detector for the ISIS and SNS. It was developed by Tremsin A.S. from the University of California, Berkeley and can reach a sub-15 μm spatial resolution for both thermal and cold neutrons [10] and a sub-microsecond timing resolution [11]. It is thus superior to the detectors applied to the JSNS in terms of spatial resolution.

The CSNS is a high-flux pulsed neutron source (25 Hz) that is mainly applied to neutron scattering, imaging, and other kinds of neutron science research [12]. The ERNI is currently being built at the CSNS with the goal of characterizing and analyzing the 3D distribution of the microstructures, defects, stresses and even dynamic processes of materials and devices. For the ERNI, a detector needs to have a microsecond-scale timing resolution and a spatial resolution of less than 100 μm at the same time. The timing resolution enables the detector to distinguish neutron energy by using the TOF approach and the high spatial resolution can be used to analyze the internal microstructure of samples in a number of applications. The TPX3Cam is a recently developed camera with a high timing resolution that is different from a CCD camera as it is based on the Timepix3 chip. It can record the coordinates (x,y), the time of arrival(TOA), with a granularity of 1.56ns, and time over threshold (TOT) with a granularity of 25 ns for the pixels fired almost at the same time [13]. The energy resolved neutron imaging detector can be realized by using the standard CCD camera-based neutron imaging detector, in which the TPX3Cam replaces the CCD camera without a timing resolution. However, the image intensifier needs to be coupled to the TPX3Cam in order to gain single photon sensitivity and overcome the high threshold of timepix3 chip. Compared with the MCP neutron counting detector, the TPX3Cam is protected from direct radiation damage because a reflective mirror is used to deflect the scintillation light to the direction vertical to that of the neutron beam. A high timing resolution can be realized considering the time precision of the TPX3Cam and the high spatial resolution can be obtained by the centroiding alrorithm [14][15].

This paper develops the prototype of an energy resolved neutron imaging detector based on TPX3Cam. Proof of principle experiments were performed in the BL20 at the CSNS to evaluate its performance and the results show that it is a viable option for use of the ERNI at the CSNS.

## 2. Experimental setup

A schematic diagram of the energy resolved neutron imaging detector is shown in Fig. 1. At the CSNS, a pulsed neutron beam was produced by the spallation reaction, with a repetition frequency of 25 Hz. The signal T0 from the accelerator represents the time of generation of the pulsed neutron beam. It was used as the trigger for TPX3Cam and was recorded for the calculation of the TOF. The pulsed neutrons transmitted through the sample and the transmitted neutrons were converted into light through the scintillation screen. The Timepix3 chip of TPX3Cam had a high detection threshold, about 1000 photons per pixel hit according to the specification in the manual of TPX3Cam. Thus, an image intensifier was needed to enhance the scintillation light induced by the neutrons. The scintillation light was deflected into the optical lens by a reflective mirror, enhanced by the image intensifier, and finally reached the Timepix3 chip. The coordinates (x,y), TOA, and TOT of each fired pixel were measured simultaneously. The TOF was calculated according to T0 and TOA, and represented energy or wavelength of the neutron.

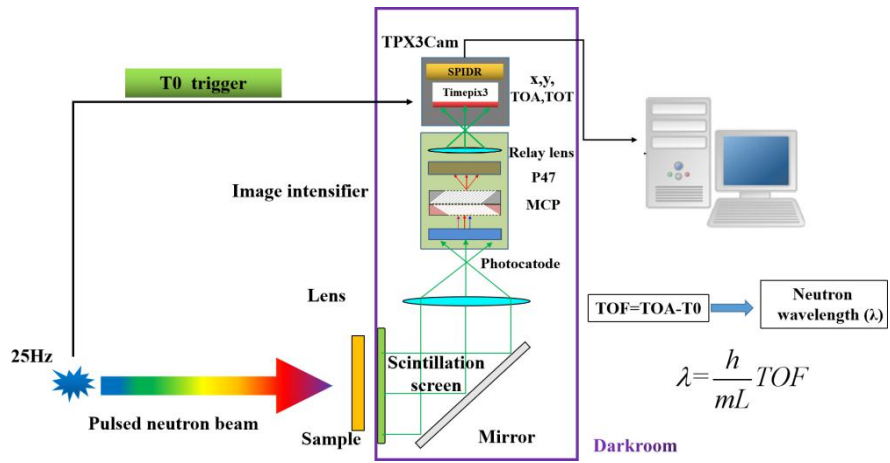

Figure 1 Schematic diagram of detector

Fig.2 shows that the detector was mainly composed of a darkroom, scintillation screen, 45° reflecting mirror, Schneider lens, 18 mm Photonis image intensifier, TPX3Cam and a movable platform. To protect the detector from radiation damage, the boron–aluminum alloy plates were installed on the inner wall of the darkroom. The scintillation screens commonly used for neutron imaging included 50 μm thick $^6$LiF/ZnS and 10 μm thick GOS ($Gd_2O_2S$ (Tb/$^6$LiF)). Compared with $^6$LiF/ZnS (Cu), GOS has better performance in terms of spatial resolution for two reasons. First, it can be made much thinner owing to higher cross-section of absorption of gadolinium. Secondly, the mean free path (MFP) of secondary particles created by GOS is much smaller than that generated by $^6$LiF/ZnS. Thus, the GOS scintillation screen was used in the spatial resolution test to limit the light spot size to about 10 μm [16][17]. However, the light output of $^6$LiF/ZnS was almost 100 times higher than that of GOS. Thus, a $^6$LiF/ZnS scintillation screen was used to reduce the experimental time when the wavelength spectrum of the neutrons was measured and Bragg edge imaging was carried out. In addition, the timing resolution of the detector was mainly affected by the scintillation screen and the TPX3Cam, and needed to be evaluated. The light decay times of the $^6$LiF/ZnS and GOS scintillation screen were on the microsecond scale[18-21]. CSNS has a high neutron flux, the detector is necessary to have a good timing resolution. The timing resolution of this detector is the minimum time interval required to separate two subsequent neutron events and it is mainly affected by the scintillation screen and the dead time of single pixel for Timepix3 chip. To analyze the effect of scintillation screen on this detector's timing resolution, the maximum time difference of fired pixels for a single neutron event can be obtained by calculating the maximum difference of measured TOA for these pixels. The results showed the maximum time difference was within sub-microsecond when $^6$LiF/ZnS or GOS scintillation screen was used. This is due to the fact that the threshold is high enough for Timepix3 chip to cut out the photons from longer tail of the decay curve. In addition, when two neutrons hit on the same position of the scintillation screen within 1μs, they are not able to be distinguished because the dead time of single pixel was 1μs for Timepix3 chip. So, the detector has a timing resolution of microsecond scale. Considering the time fluctuation caused by the moderator is on the order of microseconds, the detector can meet the requirement at CSNS.

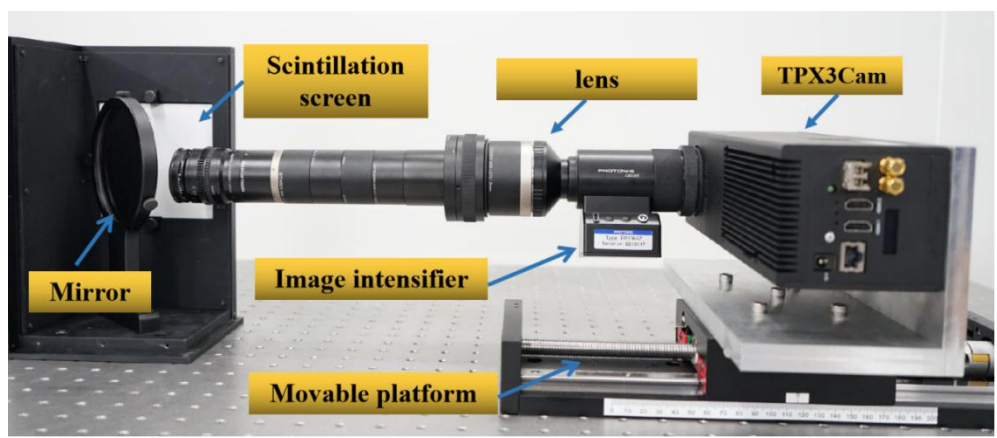

Figure 2 Setup of the detector

The peak wavelengths of the scintillation light were 530 nm for $^6$LiF/ZnS and 549 nm for GOS. Thus, the Hi-QE Blue photocathode was chosen for the image intensifier because it is compatible with the scintillation light wavelength. Meanwhile, the double MCPs were selected to increase the gain to about 90, 000 ph/ph and a fast P47 phosphor screen was chosen, with fast rise and decay times of 7 ns and 100 ns, respectively. In the measurements, 10 Gb optic fiber Ethernet was used for the TPX3Cam, and the maximum counting rate was 80 Mhits/s.

## 3. Wavelength spectrum using the TOF approach

To verify the capability of the detector in terms of the timing resolution, the wavelength spectrum was firstly measured at the BL20 with a decoupled poisoned hydrogen moderator using the TOF approach. The T0 signal needed to be connected to TPX3Cam, and the trigger mode of TPX3Cam was set to PEXSTART_TIMERSTOP, which means that the rising edge opened the shutter and it closed after the duration of exposure expired. The TOF spectrum of the empty beam was measured in the BL20 at the CSNS, and was used to obtain the neutron wavelength spectrum. It was calibrated according to the detection efficiency of the scintillation screen for neutrons of different wavelengths. The calibrated spectrum of the wavelengths of the neutrons was compared with that measured by an $^3$He beam monitor (ORDELA 4562N) [22] under the condition that these two spectra were normalized to the largest count. The results in Fig. 3 show that the spectral shape measured by the TPX3Cam detector was almost identical to that measured by the beam monitor.

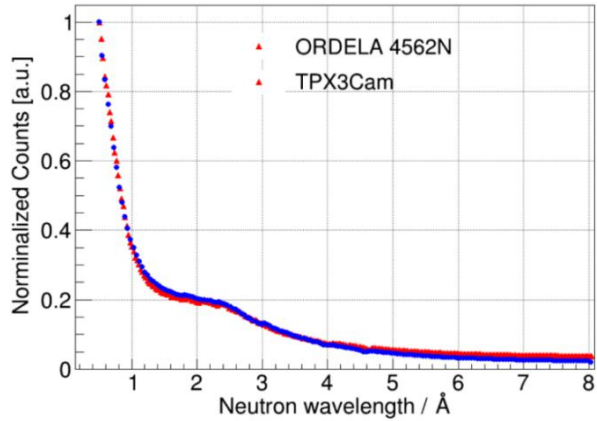

Figure 3 Neutron wavelength spectrum of CSNS BL20

## 4. Spatial resolution

Spatial resolution was defined as the minimum distance between light spots that can be distinguished. A high spatial resolution is a desirable feature of a neutron imaging detector for observing fine details of an object. For the traditional CCD camera, the pixel size is usually less than 10 μm, or between 10 μm and 20 μm. The TPX3Cam has a larger pixel size of 55 μm, but the centroiding algorithm can be used to improve its spatial resolution because the TOA, TOT, and position coordinates could be independently measured for all fired pixels. To test the centroiding algorithm, experiments on spatial resolution were carried out, including an optical test and a neutron beam test. There were two reasons for carrying out the optical spatial resolution experiments. First, the intrinsic spatial could be gotten by the optical test. Secondly, the experimental time allocated to the authors for the BL20 was short. The optical and neutron spatial resolutions were obtained when the isotopic light source which is a tritium tube was used, and the neutron beam experiment was carried out respectively. For the isotopic light source, the principle of luminescence is that β decay of the tritium gas will occur, the produced electrons excite the phosphor powder coated on the inner wall of tritium tube, and the light will be produced. The reason why the tritium tube was used is that the intensity of the produced light is very stable because the half-life of tritium is 12.43 years, and it is also very weak and suitable for the intensifier.

### 4.1 Centroiding algorithm

In the optical and neutron beam tests, the lens and image intensifier enlarged the light spot from the isotopic light source or the scintillation light. Multiple pixels were fired within sub-microseconds for a given light spot, from either the isotopic light source or a neutron event. This can lead to inaccuracies in determining the position of light or the neutron, and degrades the spatial resolution. However, the centroid of the light spot could be reconstructed, and the spatial resolution would be improved using the centroiding algorithm, which included clustering and reconstruction. In the TPX3Cam data, the TOA, TOT, and the coordinates (x,y) were recorded for every fired pixel. To implement the algorithm, all the fired pixels were sliced according to the T0 trigger signal, and pixels in one period of T0 were regarded as a processed unit for clustering and reconstruction. The principle of clustering was to determine whether any pair of fired pixels were correlated in time and space or not using the time of arrival and position coordinates. The two fired pixels were assigned the same cluster ID if they were adjacent to each other and the interval between them was less than one microsecond. Once all the pixels in a T0 period had been clustered, reconstruction would be done, whereby the centroids were calculated according to the coordinates (x,y) and TOT of the fired pixels with the same cluster ID. The centroids were regarded as the centers of initial light spots or the positions of neutron events.

4.2 Optical spatial resolution

The optical resolution tests were firstly carried out in the darkroom using the isotopic light source on the USAF (US Air Force) test patterns before the neutron imaging experiment to verify the validation of the centroiding algorithm and obtain the intrinsic spatial resolution. The tests were performed at an optical magnification of 2.6. To eliminate the non-uniformities of the light source intensity, the test patterns image was normalized by the equivalent image with no USAF test pattern present in the light source. The results of optical imaging are shown in Fig. 4 (a) and 4 (b). To analyze the spatial resolution, all the elements of group 4 were projected into the vertical direction (shown in Fig. 4 (c)). Element 4 of group 4 was resolved corresponding to 23 lp/mm (44 μm resolution). To improve the spatial resolution further, the centroiding algorithm was used to process the results of optical imaging of the test patterns, and these results are shown in Fig. 5 (a) and Fig. 5 (b), where the resolution was significantly improved. Fig. 5 (c) shows that the element 5 of group 5 was resolved and it corresponded to 57 lp/mm (18 μm resolution).

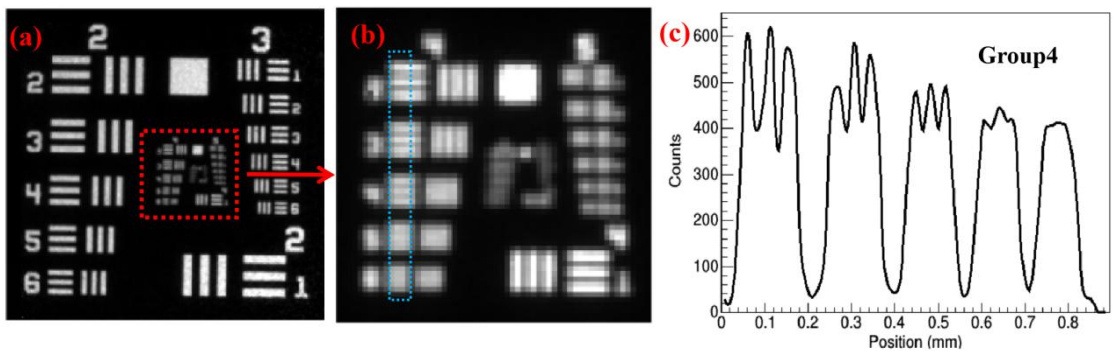

Figure 4 (a–b) Optical imaging of USAF test pattern before the centroiding algorithm (c) Projections of the Groups 4 before the centroiding algorithm, Element 4 of Group 4 is resolved corresponding to 23 lp/mm (44 μm resolution)

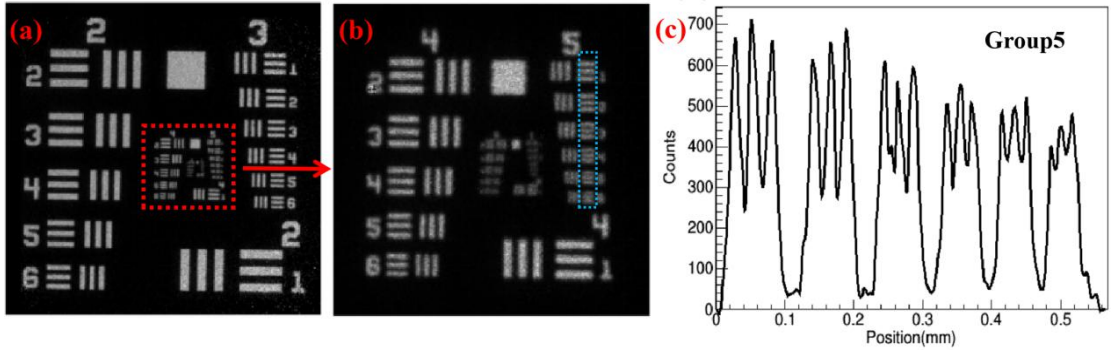

Figure 5 (a–b) Optical imaging of USAF test pattern after the centroiding algorithm (c) Projections of the Groups 5 after the centroiding algorithm. Element 6 of Group 5 is resolved corresponding to 57 lp/mm (18 μm resolution)

4.3 Neutron spatial resolution

The detector had a field of view of only 5.4 mm×5.4 mm at an optical magnification of 2.6. Thus, a laser was used to ensure that the neutron beam was vertically incident at the center of the incidence window. To increase the collimation ratio and reduce divergence effects, the detector was placed 5 m in front of the beam exit in the BL20. A Siemens Star test object, a common neutronic test pattern, was placed as close to the incidence windows of the detector as possible. The spatial resolution of neutron imaging was mainly affected by the neutron beam flux and the light spot size of the scintillator. In addition, the neutron beam was in the presence of a strong gamma radiation synchronous with T0 signal and fast neutrons which would decrease the image quality. However, they were removed by rejecting the events whose TOF was shorter than 5 ms.

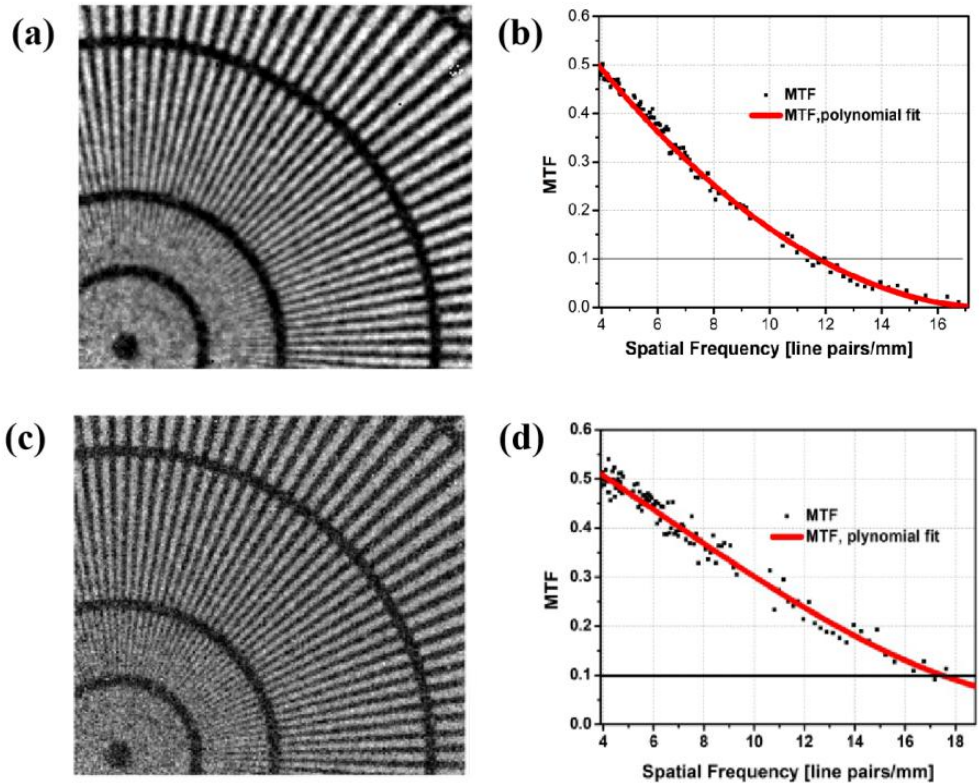

Figure 6 (a) Neutron imaging of Siemens-Star test object before the centroiding algorithm (b) MTF curve before the centroiding algorithm (c) Neutron imaging of Siemens-Star test object after the centroiding algorithm (d) MTF curve after centroiding algorithm.

The neutron images of the Siemens Star test object are shown in Fig. 6 (a), which shows that the resolution was between 50 μm and 100 μm. There are 128 line pairs in the Simens Star test object and the density of line pairs decreases with the increase of the radius from the center to the boundary of the test object. The light intensity variation of the line pairs at the every radius of the test object can be fitted by the sine function. The corresponding modulation transfer function (MTF) of different lines pair densities can be obtained by dividing the amplitude the fitted sine function by the bias level of the fitted sine function. To analyze the spatial resolution quantitatively, the MTF curve was calculated (shown in Fig. 6 (b)). When the MTF was equal to 10%, the spatial resolution was 12 lp/mm (84 μm). Fig. 6 (c) shows the result of Fig. 6 (a) after the centroiding algorithm. It is clear that the spatial resolution of this image was better than that before being processed by the centroiding algorithm. The MTF was calculated and it is shown in Fig. 6 (d). When MTF was equal to 10%, the spatial resolution of neutron imaging was 18 lp/mm (57 μm), and was worse than the optical spatial resolution. This can be attributed to two reasons. First, the neutron flux was only $10^6$ n/cm$^2$·s, and needed to be increased to improve the contrast ratio and spatial resolution of the neutron image. Secondly, the intrinsic spatial resolution of the GOS scintillation screen was 25 μm and it was a cause for the degradation of neutron spatial resolution compared to optical spatial resolution.

## 5. Neutron imaging of the total and sliced wavelengths for a CSNS Cd object

The peak cross-section of $^{113}$Cd, constituting 12.2% of naturally occurring Cd, was ~20,000 barns for thermal neutrons [23]. There was a strong energy/wavelength dependence in the transmission of the neutrons through cadmium. A CSNS pattern object of 35 × 60mm was manufactured by using a Cd plate to validate the neutron imaging of the sliced wavelengths. On the whole, the cross-section of neutrons through cadmium increased with the wavelength in the range of 0–4 Å. The neutron images of total and sliced wavelengths are shown in Fig.7 from which it was clear that the image contrast depended on the chosen neutron energy/wavelength, and gradually enhanced with the wavelength for the CSNS pattern object.

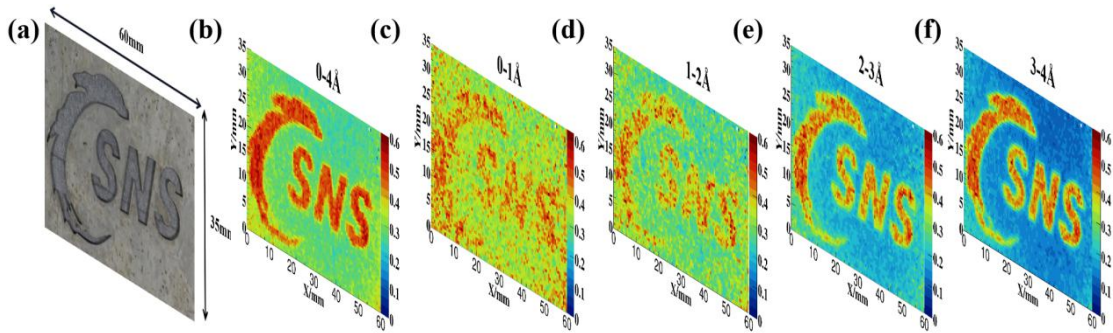

Figure 7 (a) CSNS Cd object, (b–f) Neutron images of CSNS Cd object at 0–4 Å, 0–1 Å, 1–2

Å, 2–3 Å and 3–4 Å respectively

## 6. Bragg edge imaging of a steel sample

Neutrons of different wavelengths varied in transmitted intensity through the polycrystalline samples due to diffraction. This led to a characteristic Bragg edge transmission spectrum for them. A steel sample was placed on the incidence window as shown in Fig. 8(a). It was a homogeneous polycrystalline samples. The results of neutron imaging of the sample are shown in Fig. 8(b).

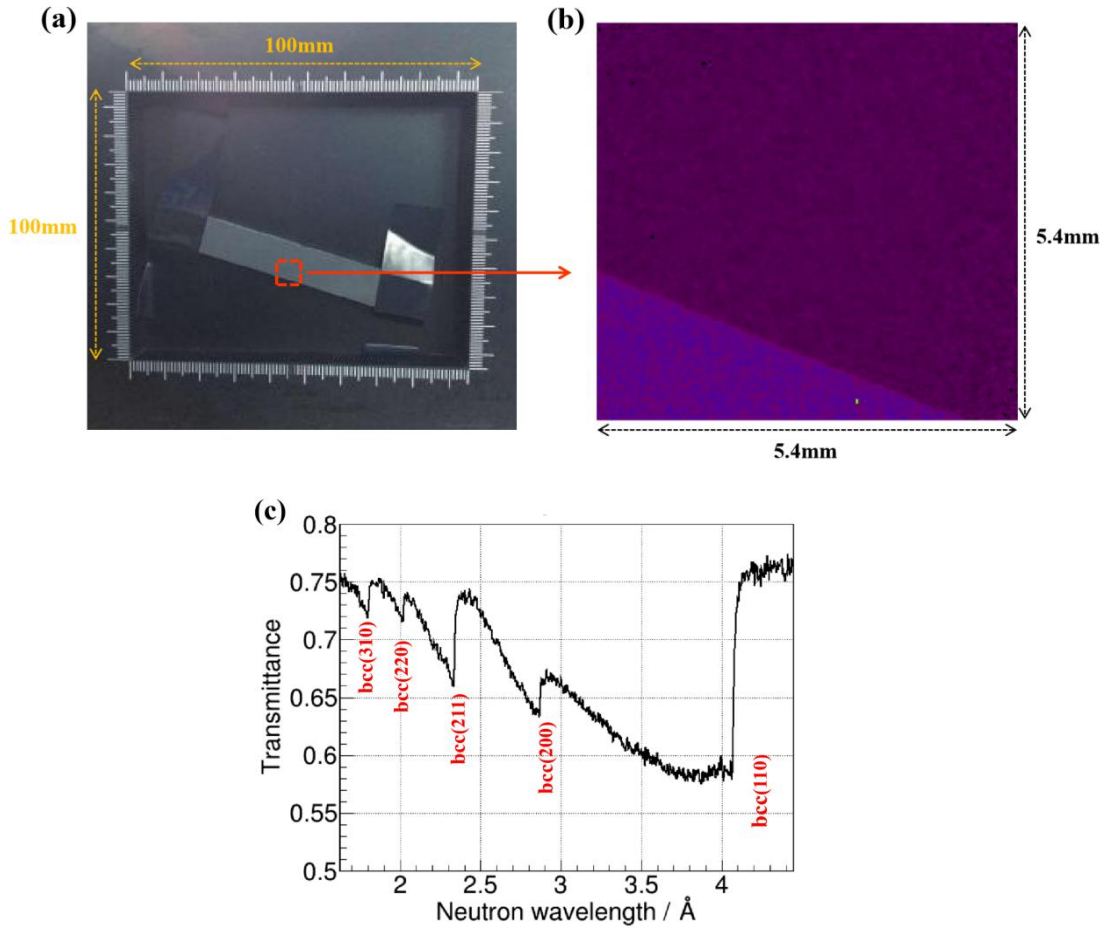

Figure 8 (a) Steel sample on incidence window (b) Neutron image of the steel sample

(c) The Bragg edge transmission spectrum

In the TOF spectrum, the maximum time of flight was 40 ms and the number of bins was set to 3000. Thus, the width of each bin was 13 μs. The wavelength spectra of the empty neutron beam and the transmitted neutrons were obtained according to the measured TOF spectrum. Based on these two spectra, the transmission wavelength spectrum of steel sample was calculated. The result is shown in Fig. 8 (c), where the corresponding lattice planes are marked. The five measured bragg edges were 4.04, 2.86, 2.32, 2.02 and 1.80 Å. The results were almost in accordance with those in Refs. [24-27]. This demonstrates that the detector can be used for the Bragg edge imaging.

## 7. Conclusion and outlook

This study constructed an energy resolved neutron imaging detector based on the TPX3Cam. The results verified the high timing resolution and spatial resolutions of this detector. It can be applied as an instrument detector for ERNI. In future work of the area, the experiments should be performed in the SANS (Small Angle Neutron Scattering Instrument) which has a higher flux than those of BL20, to further improve the spatial resolution neutron imaging. A GOS scintillation screen with a higher spatial resolution and an optical lens with a higher magnification than that employed here should be used.

## 5. Acknowledgments

This work was supported by the National Key R&D Program of China (Grant No. 2017YFA0403702), the National Natural Science Foundation of China (Grant No. U1832119, 11635012 and 11775243), Youth Innovation Promotion Association CAS, and Guangdong Basic and Applied Basic Research Foundation (Grant No. 2019A1515110217).